\documentclass[%
 reprint,
 superscriptaddress,
 amsmath,amssymb,
 aps,
 onecolumn
]{revtex4-2}

\usepackage{graphicx}
\usepackage{dcolumn}
\usepackage{color}
\usepackage{bm}
\usepackage{hyperref}
\usepackage{booktabs}

\makeatletter
\def\@author@present@script#1#2#3{%
  \begingroup
    \gdef\comma@space{\textsuperscript{}}
    \doauthor{#2}{#3}{\@affil@present@script}%
  \endgroup
  \advance\@tempcnta\m@ne
}
\makeatother

\begin{document}

\title{Rheologically tuned diffusion modulates quorum sensing in \textit{Vibrio fischeri}}

\author{Chunhe Li}
\affiliation{Global College, Shanghai Jiao Tong University, Shanghai 200240, People’s Republic of China}
\author{Zixiang Lin}
\affiliation{Global College, Shanghai Jiao Tong University, Shanghai 200240, People’s Republic of China}
\author{Hongyi Bian}
\affiliation{Global College, Shanghai Jiao Tong University, Shanghai 200240, People’s Republic of China}
\author{Anqi Li}
\affiliation{Global College, Shanghai Jiao Tong University, Shanghai 200240, People’s Republic of China}
\author{Hongyi Xin}
\affiliation{Global Institute of Future Technology, Shanghai Jiao Tong University, Shanghai 200240, People’s Republic of China}
\author{Zijie Qu}
\thanks{Contact author: \href{malto:zijie.qu@sjtu.edu.cn}{zijie.qu@sjtu.edu.cn}}
\affiliation{Global College, Shanghai Jiao Tong University, Shanghai 200240, People’s Republic of China}

\begin{abstract}
Understanding how the physical properties of a fluid influence bacterial behavior is essential for explaining how microorganisms interact with their environment and with animal hosts. Here, we examine how changes in fluid viscosity and rheological properties affect the locomotion of the marine bacterium \textit{Vibrio fischeri} and its ability to produce luminescence through cell-cell communication. We track the three-dimensional motion of single cells in well-defined fluids with different physical properties and measure the luminescence emitted by cell populations. We find that fluids with higher viscosity cause \textit{V.~fischeri} to spend more time in a slower, turning-focused swimming mode, which reduces how effectively cells spread out and encounter the chemical signals required to activate luminescence. As a result, luminescence first increases and then decreases in Newtonian fluids, but decreases monotonically in fluids that exhibit non-Newtonian rheological behavior. Computer simulations based on our measurements confirm that the ability of cells to explore their surroundings plays a central role in determining when and how strongly they communicate. These findings reveal a direct link between the physical environment, bacterial movement, and collective behavior, and offer insight into how microorganisms adapt to complex fluid habitats, including those found inside animal hosts.
\end{abstract}

\maketitle

\section{Introduction}

Microorganisms inhabit physically diverse liquid environments in which the physical properties of the surrounding fluid strongly influence their ability to navigate, sense, and communicate across multiple spatial and temporal scales~\cite{Chaithanya_et_al_2025,Elfring_Lauga_2014,Lauga_2016,Quan_et_al_2025}. Viscosity, elasticity, and shear-thinning behavior shape not only the hydrodynamic forces experienced by swimming cells but also the transport~\cite{Wong_et_al_2023,Winkle_et_al_2023}, accumulation~\cite{Scheidweiler_et_al_2024}, and perception of signaling molecules that coordinate collective behaviors~\cite{Cornforth_et_al_2014,Boyer_et_al_2009}. Such physical constraints are ubiquitous in natural habitats, including marine snow~\cite{Gram_et_al_2002,BarZeev_Dang_2016}, host mucus layers~\cite{Lai_et_al_2009,Mrokowska_et_al_2022}, and polymer-rich biological secretions, where fluids frequently deviate from Newtonian behavior and exhibit complex rheological responses~\cite{Passow_2001,The_biology_of_mucus_2023}. Understanding how environmental rheology regulates cellular locomotion and chemical communication is therefore essential for linking microscopic motility to ecological and host-associated functions.

\begin{figure}
\centering
\includegraphics[width=0.8\linewidth]{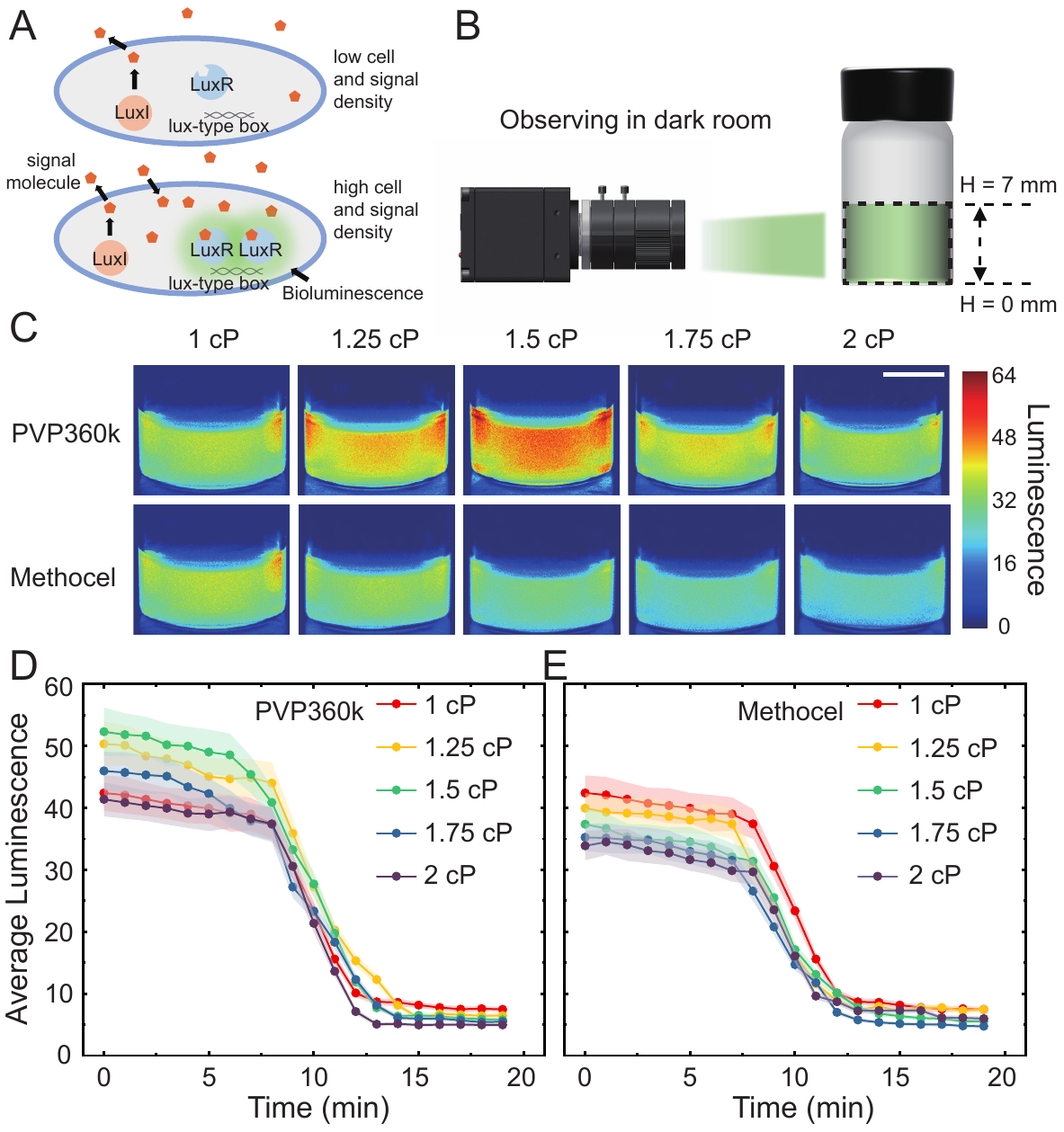}
\caption{Viscosity-dependent modulation of quorum-sensing–mediated bioluminescence in \textit{V.~fischeri}. (A) The schematic diagram illustrating QS-activated bioluminescence mechanism of \textit{V.~fischeri}. (B) Experimental setup diagram for observing the variation of bioluminescence intensity with the rheological properties of the solution. (C) Representative false-color images of bioluminescence intensity for bacterial suspensions at fixed cell concentration in motility buffers containing PVP360k (Newtonian fluid, top row) and Methocel (bottom row) at different viscosities. Color scale indicates luminescence intensity. Temporal evolution of the spatially averaged luminescence intensity in PVP360k (D) and Methocel (E) containing motility buffer for viscosities ranging from 1 to 2~cP. {The solid points represent the average intensity, and the shaded areas indicate the standard deviation across five independent experiments.} The mean luminescence intensity exhibits a nonmonotonic dependence on viscosity, with a maximum at intermediate viscosity. Scale bar: $5~mm$. \label{brightness}}
\end{figure}

A compelling illustration of the ecological relevance of fluid--cell interactions is the binary symbiosis between \textit{Vibrio fischeri} and the Hawaiian bobtail squid, \textit{Euprymna scolopes}~\cite{Nyholm_et_al_2000,Nawroth_et_al_2017,Nyholm_McFallNgai_2003}. {Newly hatched squid depend on motile \textit{V.~fischeri} cells to traverse a heterogeneous, mucus-rich landscape and reach the light organ, where successful colonization requires cells to accumulate, persist, and communicate within spatially confined crypts~\cite{nyholm2021lasting,Gundlach_et_al_2022,Chavez-Dozal_et_al_2012} (see Fig.~S1)}. At the cellular level, the QS-activated bioluminescence mechanism of \textit{V.~fischeri} is illustrated as shown in the Fig.~\ref{brightness}A. At low cell and signal density, LuxR remains inactive; accumulation of autoinducer molecules at high density activates LuxR binding to the lux-type box, triggering bioluminescence. Within specific microenvironments, the buildup of autoinducer molecules activates quorum sensing (QS), triggering bioluminescence that enables the squid’s nocturnal counterillumination~\cite{Engebrecht_Silverman_1989,Fuqua_Stevens_Pearson_1994}. This iconic symbiosis thus exemplifies how bacterial motility, environmental physical properties, and QS activation are tightly coupled, yet the physical principles governing this coupling remain incompletely understood.

Previous studies have established that bacterial motility is highly sensitive to fluid rheology. In Newtonian fluids, increased viscosity generally reduces swimming speed by enhancing hydrodynamic drag~\cite{Qin_et_al_2015,Keller_1974,Lauga_Powers_2009,Qu_et_al_2018}. In contrast, dilute polymer solutions can produce non-intuitive responses: for example, \textit{E. coli} has been reported to swim faster in certain shear-thinning polymer solutions, in contrast to its behavior in Newtonian fluids~\cite{Qu2020ShearThinningViscoelastic,Patteson_et_al_2015,Zottl_Yeomans_2019}. These findings demonstrate that the impact of complex fluids on bacterial locomotion is species-specific and depends on cell morphology, flagellar architecture, and motor regulation. Despite extensive work on the QS circuitry and ecological role of \textit{V.~fischeri}, how its motility responds to non-Newtonian, mucus-like environments---and how such responses feed back into QS-mediated behaviors---remains largely unexplored.

A distinctive feature of \textit{V.~fischeri} is its polymorphic swimming behavior, characterized by run--reverse and run--pause--reverse modes enabled by a flagellar wrapping mechanism. {In the run--reverse mode, reversal of flagellar motor rotation switches the flagellar bundle between the push and pull configurations, causing the cell to abruptly reverse its swimming direction and proceed approximately along its previous path. In the run--pause--reverse mode, the flagellar bundle transiently wraps around the cell body during the transition between two oppositely directed runs. Because propulsion is substantially reduced in the wrap configuration, this intermediate state appears as a pause in the cell trajectory. The cell subsequently resumes straight swimming in the opposite or a reoriented direction after the flagellar bundle unwraps and returns to an extended configuration~\cite{Zhuang_Tseng_Lo_2024}.} These swimming modes generate large reorientation angles and strongly influence how efficiently cells explore confined or structured spaces~\cite{Lynch_et_al_2022,Kinosita_et_al_2018,Stocker_et_al_2008,Tian_et_al_2022}. Because QS activation depends on the accumulation of signal molecules over time, the long-time effective diffusion of a bacterial population, rather than instantaneous swimming speed alone, is likely to govern the onset and spatial organization of bioluminescence~\cite{Patteson_et_al_2015,Villa-Torrealba_et_al_2020,Guccione_et_al_2017}. However, how this diffusion-like behavior emerges from viscosity-dependent swimming dynamics, particularly under viscoelastic stresses, has not been experimentally quantified.

Here, we investigate how fluid rheological properties regulate both swimming behavior and QS-mediated bioluminescence in \textit{V.~fischeri}. {Using Newtonian solutions of poly(vinylpyrrolidone) (PVP; $M_w \approx 360$ kDa, denoted as PVP360k) and methylcellulose (Methocel) solutions exhibiting weak shear-dependent rheological behavior at the concentrations used here,} we combine real-time three-dimensional tracking of individual cells with population-level measurements of QS-driven luminescence. {The rheological properties of the PVP360k and Methocel solutions are shown in Fig.~S2 of the Supplemental Material~\cite{SM}.} By quantifying motility-mode fractions, run durations, reorientation statistics, and swimming speeds across a range of viscosities, we derive the effective diffusion coefficient of \textit{V.~fischeri} and directly link it to QS activation. Complementary simulations coupling bacterial motion with viscosity-dependent molecular diffusion reproduce the observed luminescence trends, revealing how fluid rheology modulates collective bacterial signaling through its impact on long-time transport statistics. Together, our results identify effective diffusion as a physical bridge between single-cell motility and QS-controlled symbiotic function.

\section{Results and Discussion}

\subsection{Effect of Liquid Viscosity on QS-Mediated Bioluminescence}

To investigate how fluid rheological properties influence quorum-sensing (QS)--regulated bioluminescence in \textit{V.~fischeri}, we systematically tune the physical properties of the surrounding medium using PVP360k, which maintains the medium as a Newtonian fluid, and  {Methocel, which introduces shear-thinning and viscoelastic characteristics at sufficient polymer concentrations~\cite{Qu2020ShearThinningViscoelastic}}. {At the relatively low Methocel concentrations used in the present experiments, repeated bulk rheological measurements reveal only a weak decrease in viscosity with increasing shear rate (Fig. S3).} We then examine changes in bioluminescence intensity for bacterial suspensions at fixed cell concentration under different viscosities. Cells cultured for more than 24 h are harvested, centrifuged, and resuspended in motility buffer at a fixed cell concentration (details shown in the SI). As shown in Fig.~\ref{brightness}B, the vial is placed in a dark chamber, and a color camera records the bioluminescence intensity and duration by continuous imaging.

We find that the viscosity-dependent response of bioluminescence differs qualitatively between the PVP360k- and Methocel-containing media (Fig.~\ref{brightness}C). For bacteria suspended in motility buffer containing PVP360k, the bioluminescence intensity increases and then decreases with increasing viscosity, reaching a maximum at a viscosity of 1.5~cP. In contrast, for bacteria in motility buffer containing Methocel, the mean luminescence decreases monotonically with viscosity. Fig.~\ref{brightness}D and~\ref{brightness}E show the temporal evolution of the mean intensity in the luminescent region. By shaking the sample evenly and placing it in the observation chamber, this is recorded as $t = 0$. The average luminescence intensity depends on both the viscosity and the rheological properties of the motility buffer, whereas the overall duration of bioluminescence is independent of these factors. In both PVP360k- and Methocel-containing buffers, bioluminescence decays slowly during the first 7-8~min of observation and drops to its minimum level after approximately 13~min. {Because molecular oxygen is required for the luciferase-catalyzed bioluminescence reaction in \textit{V. fischeri}~\cite{Makemson1986Luciferase,Bourgois2001Kinetics}, we infer that this overall decay primarily results from the progressive depletion of dissolved oxygen in the medium. As oxygen becomes limiting, the bioluminescence intensity decreases more rapidly.} This distinct viscosity-dependent modulation of luminescence intensity motivates us to further examine the swimming behavior of individual \textit{V.~fischeri} cell.

\subsection{Effect of Complex Fluids on the Swimming Speed of \textit{V.~fischeri}}

\begin{figure}
\centering
\includegraphics[width=1\linewidth]{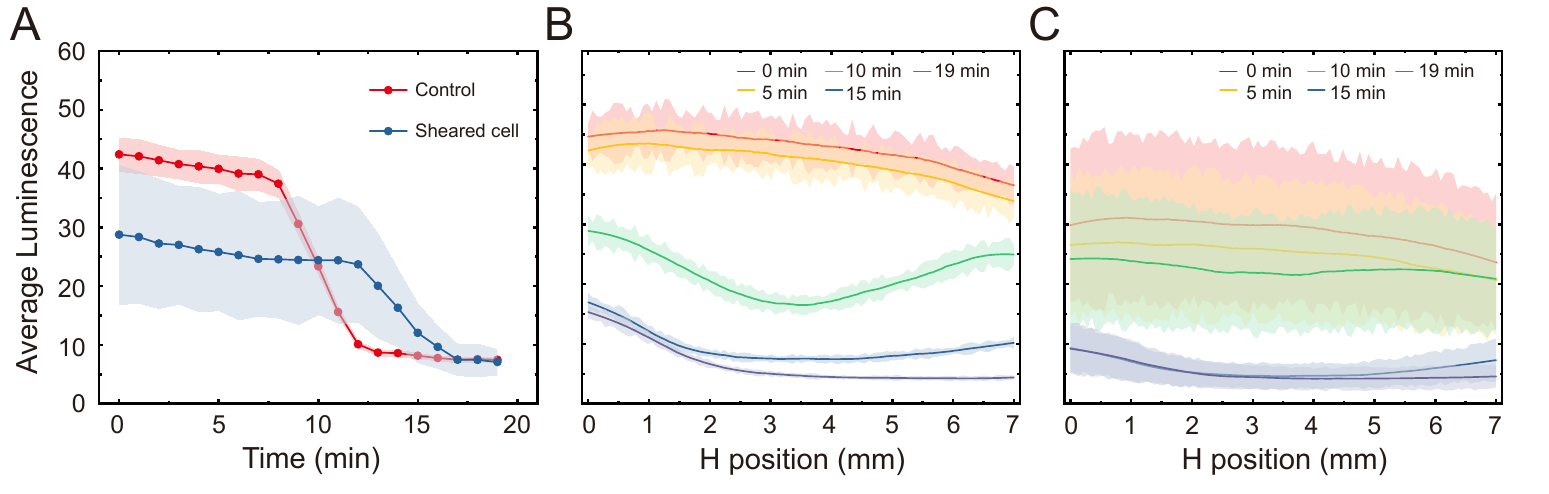}
\caption{
{ Average luminescence over time and at various $H$ positions for control and sheared cells in polymer-free TMN motility buffer. Shaded regions indicate the standard deviation across independent experiments. (A) Time-dependent average luminescence for control and sheared cells. (B) Average luminescence across different $H$ positions at various time points, including 0, 5, 10, and 15 min. (C) Similar analysis for sheared cells at the same time points, showing the position-dependent luminescence profiles during the decay process.}\label{shear}}
\end{figure}

First, we subjected \textit{V.~fischeri} cells to flagellar shearing to investigate whether the loss of motility affects bioluminescence during quorum sensing (QS). The results, shown in Fig.~\ref{shear}A, reveal that the average bioluminescence intensity of the sheared cells is significantly lower than that of the control group. This suggests that motility loss negatively impacts bioluminescence during QS. Interestingly, the duration of bioluminescence differs between the groups. The sheared cells exhibit a prolonged bioluminescence duration compared to the control, gradually decreasing after approximately 12 minutes and reaching a minimal level around 17 minutes. We hypothesize that this extended bioluminescence is due to hydrodynamic attraction at the solid-liquid interface and the cells' oxygen tropism. To further explore this, we analyze the changes in bioluminescence intensity along the H direction in both the control and sheared cells during QS. Fig.~\ref{shear}B shows that the control group initially exhibits a uniform bioluminescence distribution along the H direction during QS. However, around 10 minutes, a significant deviation in brightness is observed, with stronger bioluminescence near $H=0$ mm and $H=7$ mm compared to the central region, before the overall brightness decreases to a low level. Fig.~\ref{shear}C shows that in the sheared cells, the brightness distribution along the H direction remains relatively uniform throughout the bioluminescence decay process.

\begin{figure}
\centering
\includegraphics[width=1\linewidth]{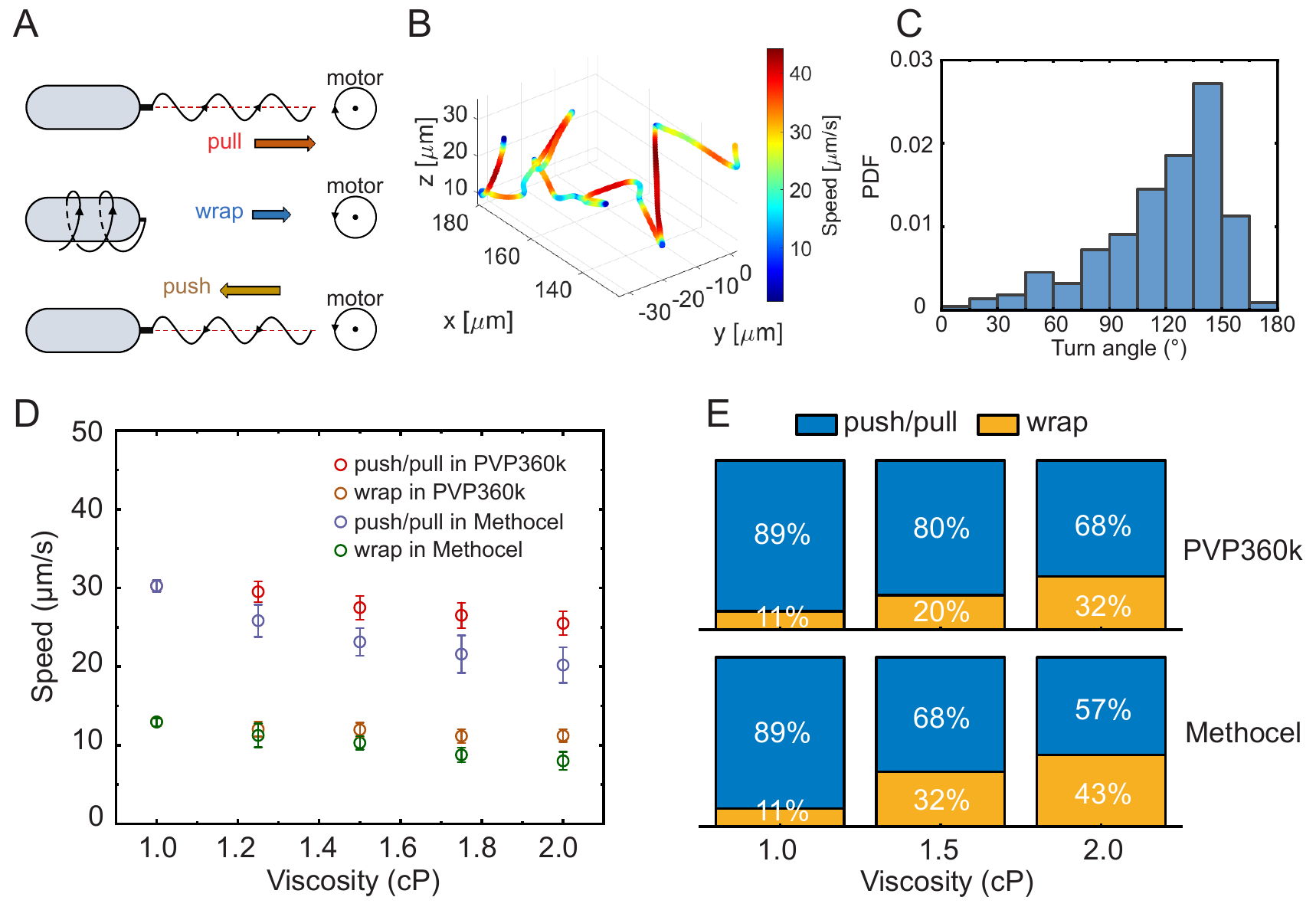}
\caption{
Characterization of swimming behavior of \textit{V. fischeri} under different viscosities and swimming modes. (A) Schematic representation of swimming modes: pull, wrap, and push. (B) 3D trajectory of \textit{V. fischeri}, with speed indicated by the color gradient. The black arrow denotes the direction of flagellar rotation. The orange, blue, and yellow arrows represent the swimming directions of bacteria in the pull, wrap, and push modes, respectively. The circular symbols with arrows on the right illustrate the corresponding motor rotation directions for the three swimming modes. (C) Probability distribution function (PDF) of turn angles, illustrating the distribution of turn angles during movement. (D) Speed of \textit{V. fischeri} in PVP360k and Methocel solutions across different viscosities, with error bars showing variability. (E) Fraction of time spent in push/pull versus wrap modes at different viscosities in PVP360k and Methocel, showing a shift in mode preference with increasing viscosity. 
\label{speed}}
\end{figure}

We measure and analyze the swimming behavior of \textit{V.~fischeri} in motility buffers with different fluid properties by tracking each individual cell for 10~s using a real-time 3D tracking microscope~\cite{Qu_et_al_2018}. To elucidate the origin of this viscosity-dependent slowdown, we further examine the motility modes of individual cells. As illustrated in Fig.~\ref{speed}A, entry into the wrap mode is accompanied by a substantial reduction in instantaneous swimming speed compared to the push and pull modes. {Fig.~\ref{speed}B shows a representative 3D trajectory of \textit{Vibrio fischeri} in TMN motility buffer without added polymer.} {The swimming modes are identified from the smoothed three-dimensional trajectories based on changes in swimming direction and instantaneous speed, as described in Section~4.5.} It is evident that the cell consistently exhibits significant deceleration and angular deflection during wrapping. We analyze the histogram of angular changes before and after wrapping in a motility buffer without any added polymer and find that the turn angle is concentrated around 140°~(Fig.~\ref{speed}C). 
We measure and analyze the swimming behavior of \textit{V.~fischeri} in motility buffers with different fluid properties by tracking each individual cell for 10~s using a real-time 3D tracking microscope~\cite{Qu_et_al_2018}. To elucidate the origin of this viscosity-dependent slowdown, we further examine the motility modes of individual cells. As illustrated in Fig.~\ref{speed}A, entry into the wrap mode is accompanied by a substantial reduction in instantaneous swimming speed compared to the push and pull modes. {Fig.~\ref{speed}B shows a representative 3D trajectory of \textit{Vibrio fischeri} in TMN motility buffer without added polymer.} It is evident that the cell consistently exhibits significant deceleration and angular deflection during wrapping. We analyze the histogram of angular changes before and after wrapping in a motility buffer without any added polymer and find that the turn angle is concentrated around 140°~(Fig.~\ref{speed}C).

In motility buffer supplemented with PVP360k, the mean swimming speed decreases monotonically with increasing viscosity, consistent with expectations for Newtonian fluids. We therefore segment the three-dimensional trajectories to quantify both the fraction of time spent in each motility mode and the corresponding mode-resolved swimming speeds across different viscosities. Fig.~\ref{speed}D shows that swimming speeds decrease in both the straight swimming modes (push and pull) and the wrap mode. {In Newtonian fluids, the observed decrease in swimming speed with increasing viscosity can be rationalized within the framework of resistive force theory (RFT), as detailed in Fig.~S3 of the Supplemental Material~\cite{SM}. The corresponding local shear rate near the rotating flagella is estimated to reach several $10^3~\mathrm{s^{-1}}$ based on the same RFT parameters.} Interestingly, for the same viscosity, swimming speeds in Methocel solutions are lower than in the Newtonian PVP360k solution, whether in the straight swimming modes (push and pull) or the wrap mode. These differences at matched bulk viscosities indicate that the bacterial motility response cannot be explained by bulk viscosity alone. In Fig.~\ref{speed}E, the probability of the wrap mode increases systematically with viscosity in both PVP360k and Methocel-containing buffers. This trend is consistent with previous studies, where increased viscous resistance promotes the occurrence of flagellar wrapping events~\cite{Zhuang_Tseng_Lo_2024}. However, in contrast to earlier studies on \textit{Escherichia coli}~\cite{Qu2020ShearThinningViscoelastic,Patteson_et_al_2015,Zottl_Yeomans_2019}, \textit{V.~fischeri} does not exhibit enhanced swimming speed with increasing viscosity in shear-thinning fluids. 

\subsection{Relationship between the Diffusion and Bioluminescence}

\begin{figure}
\centering
\includegraphics[width=1\linewidth]{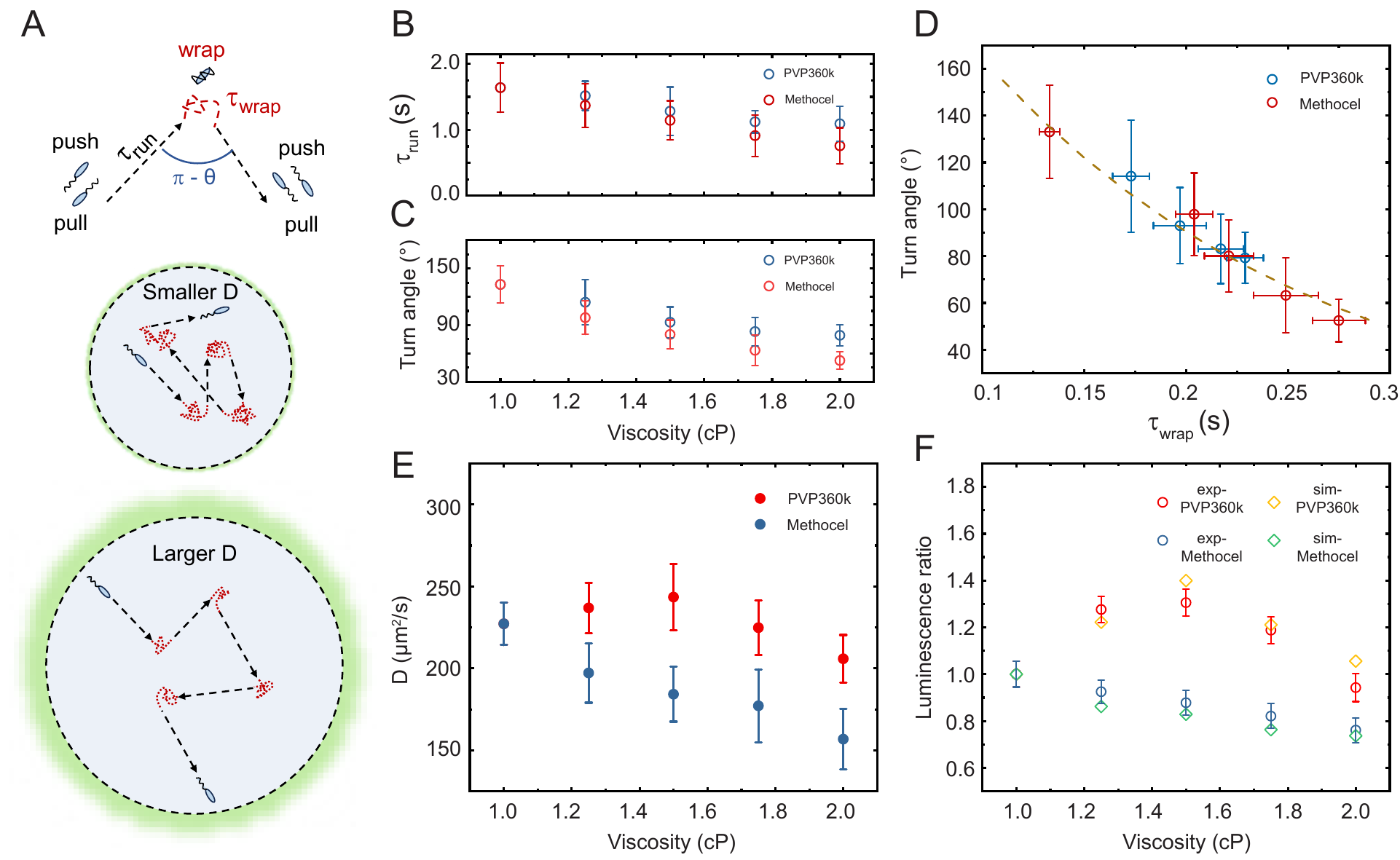}
\caption{
Motility statistics, effective diffusion, and their connection to quorum-sensing activity in \textit{V.~fischeri}. 
(A) Schematic illustration of the motility statistics used to characterize cell reorientation and environmental exploration. 
Straight swimming in push and pull modes is intermittently interrupted by wrap events of duration $\tau_{\mathrm{wrap}}$, leading to a reorientation characterized by the turning angle $\theta$ and an effective run duration $\tau_{\mathrm{run}}$. 
Representative trajectories illustrate how different motility statistics give rise to smaller or larger effective diffusion coefficients $D$. 
{(B) Mean run duration $\tau_{\mathrm{run}}$ as a function of viscosity.
(C) Mean turning angle as a function of viscosity.
(D) Mean turning angle as a function of the wrap duration $\tau_{\mathrm{wrap}}$ for cells swimming in PVP360k- and Methocel-containing motility buffers. 
Dash line represents exponential fits to the data. 
(E) Effective diffusion coefficient $D$ calculated from the measured swimming speed, run duration, and turning-angle statistics. Error bar represents the standard error.}
(F) Comparison between the normalized mean luminescence intensity measured experimentally and simulation results based on a diffusion-controlled interaction model between cells and signal molecules. \label{Diffusion}}
\end{figure}

Previous studies indicate that quorum sensing in \textit{V.~fischeri} is activated once cells encounter a sufficiently high local concentration of signal molecules, leading to the onset of bioluminescence~\cite{Schaefer_et_al_1996,Lupp_Ruby_2005}. This suggests that the efficiency with which cells explore their surroundings plays a key role in QS activation, and can be quantitatively characterized by an effective diffusion coefficient $D$~\cite{Patteson_et_al_2015,Jepson_et_al_2013}. {It should be noted that QS activation is governed by both biochemical regulation of signal production and accumulation and by the physical transport processes that determine cell–signal encounters. Here, we focus specifically on the latter contribution and do not imply that effective diffusion alone determines QS activation.} The swimming behavior of \textit{V.~fischeri} consists of straight runs at speed $v$, intermittently interrupted by reorientation events associated with flagellar wrapping and unwrapping. These reorientation events are characterized by a run duration $\tau_{\mathrm{run}}$, a wrapping time $\tau_{\mathrm{wrap}}$, and a turning angle $\theta$ (Fig.~\ref{Diffusion}A). {Histograms showing the distributions of $\tau_{\mathrm{run}}$ and $\tau_{\mathrm{wrap}}$ in PVP360k and Methocel solutions at different mass fractions are provided in Figs.~S4 and S5 of the Supplemental Material~\cite{SM}, respectively. In addition, the wrapping duration $\tau_{\mathrm{wrap}}$ is consistently much shorter than the corresponding run time $\tau_{\mathrm{run}}$ across all viscosities.} Analysis of the trajectory statistics shows that the distribution of run durations is well described by an exponential form, supporting a Poisson description of reorientation events~\cite{Taktikos_Stark_Zaburdaev_2013}.

Using the measured three-dimensional trajectories, we extract $v$, $\tau_{\mathrm{run}}$, and the turning-angle statistics at each viscosity for both motility buffers. Both the mean run duration and the mean turning angle decrease monotonically with increasing viscosity in PVP360k and Methocel solutions (Fig.~\ref{Diffusion}B and C).  The mean turning angle decreases systematically with increasing $\tau_{\mathrm{wrap}}$, and the dependence is well captured by an exponential fit for both PVP360k- and Methocel-containing buffers (Fig.~\ref{Diffusion}D), indicating that longer wrapping events lead to more persistent swimming directions~\cite{Tian_et_al_2022}. The ability of bacteria to explore the $ D = v^2 \tau_{\mathrm{run}} / 3 \left( 1 - \langle \cos \theta \rangle \right)$~\cite{Taktikos_Stark_Zaburdaev_2013}. Substituting these measured quantities into equation yields the effective diffusion coefficient $D$. As shown in Fig.~\ref{Diffusion}D, $D$ exhibits a nonmonotonic dependence on viscosity in PVP360k containing Newtonian fluids, whereas it decreases monotonically with viscosity in Methocel containing viscoelastic fluids. 

To connect these transport properties to QS activity, we quantify the mean bioluminescence intensity during the emission period for cell populations at identical concentration. The viscosity dependence of the luminescence intensity closely mirrors that of the effective diffusion coefficient for both motility buffers (Fig.~\ref{Diffusion}D and E). We employ a simplified diffusion-based model to simulate the interaction between \textit{V.~fischeri} cells and surrounding signal molecules under representative experimental conditions ($16^{\circ}\mathrm{C}$) in a cubic domain with a side length of $300~\mu\mathrm{m}$. In this model, the active motion of the cells is described by the effective diffusion coefficients measured at different viscosities in the two motility buffers, while the passive motion of signal is modeled using the viscosity-dependent diffusion coefficients of small molecules. The normalized number of signal--cell encounter events is used to quantify the luminescence intensity. As shown in Fig.~\ref{Diffusion}E, the simulations reproduce the overall experimental trends. Together, these results indicate that the “searching efficiency” of the cells, quantified by the effective diffusion coefficient, constitutes a key physical factor linking motility statistics to QS-activated bioluminescence.

\section{Conclusion}

In this work, we examine how liquid viscosity and fluid rheology regulate quorum-sensing (QS)–mediated bioluminescence in \textit{V.~fischeri}, under the hypothesis that viscosity-dependent changes in bacterial motility—rather than viscosity alone—govern the efficiency of signal encounter and accumulation. {By systematically comparing PVP360k- and Methocel-containing fluids with matched bulk viscosities,} we aim to establish a physical link between microbial transport, motility modes, and QS activation learned from experiments and modeling. Our results reveal that fluid rheology plays a decisive role in shaping QS-regulated bioluminescence. In Newtonian fluids, the luminescence intensity exhibits a nonmonotonic dependence on viscosity, whereas in viscoelastic Methocel solutions it decreases monotonically. Single-cell tracking demonstrates that increasing viscosity improves the probability of flagellar wrapping and restricts swimming speed, with a stronger reduction observed in Methocel-containing media. Unlike previously reported viscosity-enhanced motility in \textit{E.coli}, \textit{V.~fischeri} shows no speed enhancement in shear-thinning media, underscoring the importance of species-specific motility mechanisms and flagellar dynamics. Considering that this is a phenomenon worth investigating, but not the main focus of our current work, we will conduct detailed experimental and theoretical studies on the speed reduction behavior of \textit{V.~fischeri} in viscoelastic fluids in future work.

Importantly, these findings resonate with the natural symbiotic context of \textit{V.~fischeri} and the Hawaiian bobtail squid \textit{Euprymna scolopes}. During the initial colonization process, ciliary beating on the squid light-organ surface captures bacteria from seawater while simultaneously secreting mucus that is known to exhibit non-Newtonian, viscoelastic properties. Previous studies show that bacterial swimming speeds that are either too high or too low reduce capture and retention efficiency by the host~\cite{Millikan_Ruby_2002,Graf_Dunlap_Ruby_1994}. Our observation that V. fischeri experiences a pronounced motility slowdown in non-Newtonian fluids therefore aligns with this ecological constraint, suggesting that fluid rheology may act as a physical regulator that tunes bacterial transport during host-mediated selection and accumulation.

By quantifying swimming speed, turning angles, and motility-mode fractions, we derive an effective diffusion coefficient that characterizes how efficiently cells explore their environment. We show that the viscosity-dependent variation of this diffusion coefficient consistently predicts the observed changes in luminescence intensity across both Newtonian and non-Newtonian fluids. A diffusion-based interaction model between cells and signal molecules further reproduces the experimental trends, highlighting “search efficiency” as a key physical control parameter linking hydrodynamic transport to collective biochemical activation.

Several limitations of this study should be acknowledged. {The effective diffusion description isolates the physical contribution of bacterial transport to signal encounters and does not represent the full biochemical QS process. In particular, the present model does not explicitly account for time-dependent signal production and degradation, intracellular LuxI/LuxR feedback, biochemical coupling between QS activation and motility, or heterogeneity in signal production. Therefore, effective diffusion should be regarded as one physical factor modulating QS activation rather than its sole determinant.} Future work may address these limitations by coupling intracellular QS regulation to motility dynamics, probing near-field viscoelastic effects at the single-flagellum scale, and extending the framework to higher cell densities or structured environments.  Exploring how fluid rheology modulates QS in other motile bacteria may reveal further the general physical principles that govern microbial communication.

\begingroup

\section{Materials and methods}
\subsection{Cell culture}\label{V.Cell culture}
\textit{Vibrio fischeri} cells (strain CICC24882) used in the experiments are obtained from the China Industrial Microbial Culture Collection and Management Center (CICC). The cells are cultured in Marine Broth 2216E in 250~mL Erlenmeyer flasks sealed with gas-permeable membranes, which allow sufficient gas exchange while minimizing external perturbations. Cultures are maintained at $16~^\circ\mathrm{C}$. Cells are first grown in Marine Broth 2216E for approximately 24~h and subsequently diluted 1:500 into fresh media, followed by an additional 6–8~h of growth prior to experiments. At this stage, cells exhibit stable swimming behavior and high viability, ensuring consistent experimental conditions.

\subsection{Sample preparation}
After cultivation, \textit{V.~fischeri} cells are gently transferred into TMN motility buffer (Tris--HCl-based buffer containing NaCl, MgCl$_2$, and glucose) for swimming behavior observations. TMN provides a well-defined ionic and energetic environment that sustains motility while suppressing uncontrolled growth during measurements (Table~\ref{TMN_composition}).  

For experiments involving complex fluids, polyvinylpyrrolidone (PVP, $M_w \approx 360$ kDa) and/or Methocel (methylcellulose) are added to TMN at prescribed concentrations expressed as weight/volume percentages (\%~w/v), depending on the experimental objective. The PVP concentrations used in this study are 0, 0.21, 0.37, 0.50, and 0.62\%~w/v, while the Methocel concentrations are 0, 0.05, 0.10, 0.15, and 0.17\%~w/v. Following polymer addition, the solutions are mixed on an orbital shaker at 200~rpm for more than 12~h to ensure complete polymer dissolution.

\begin{table}[htbp]
\centering
\caption{Composition of TMN motility buffer (1$\times$) per liter~\cite{Zhuang_Tseng_Lo_2024}}
\label{TMN_composition}
\begin{tabular}{l c c}
\toprule
\textbf{Component} & \textbf{Chemical formula} & \textbf{Amount per 1 L} \\
\midrule
Tris--HCl buffer & -- & 50 mM \\
Sodium chloride & NaCl & 17.5 g \\
Magnesium chloride (anhydrous) & MgCl$_2$ & 0.48 g \\
D-Glucose & C$_6$H$_{12}$O$_6$ & 0.90 g \\
\midrule
Deionized water & -- & to 1 L \\
\bottomrule
\end{tabular}
\end{table}

\subsection{Bioluminescence imaging setup}

Quorum-sensing–induced bioluminescence is monitored using an industrial CMOS camera (model MV-CU120-10UC) operated in continuous acquisition mode. Image acquisition and camera control are performed using the manufacturer-provided machine vision software (HIKROBOT MVS, version~4.5.1). A fixed-focus lens (MVL-HF0624M-10MP) is mounted to provide a stable field of view and high light-collection efficiency.

All measurements are conducted in a light-tight darkroom to eliminate background illumination. The imaging system is maintained at a constant temperature of $16~^\circ\mathrm{C}$ throughout the experiment to ensure stable physiological conditions. Time-lapse images are recorded at a temporal resolution of one frame per minute, enabling continuous tracking of the onset and temporal evolution of bacterial bioluminescence.

The cell concentration is then quantified using optical density measurements at 600~nm (OD$_{600}$). After concentration calibration, the samples are placed in a dark environment, where the bioluminescence emitted by the bacterial population is recorded. Bioluminescence is monitored by time-lapse imaging, and the recorded images are converted to 8-bit grayscale format, with pixel intensities ranging from 0 to 255. The average grayscale intensity is used as a quantitative measure of bioluminescence output, allowing consistent comparison across different fluid conditions. The results are summarized in Fig.~S6. The bioluminescence intensity increases monotonically with cell concentration, demonstrating a clear positive correlation between QS-activated light emission and population density. In contrast, the duration of the luminescence signal decreases as the cell concentration increases. This inverse relationship is likely due to rapid depletion of dissolved oxygen at high cell densities, which limits the sustainability of the bioluminescent reaction.

\subsection{Flagellar shearing}

Flagella are mechanically removed from \textit{V.~fischeri} cells by repeated aspiration and dispensing. Briefly, $100~\mu\mathrm{L}$ of the bacterial suspension is aspirated using a pipette (Eppendorf, 10--200~$\mu\mathrm{L}$) and then rapidly dispensed from the pipette tip. This aspiration--dispensing cycle is repeated 50 times to shear the flagella from the bacterial cells. Comparisons of bacterial cells before and after flagellar shearing are provided in SI (Movies~S7 and S8). {Because flagellar removal was performed by repeated manual aspiration and dispensing, small variations in the applied mechanical shear may lead to batch-to-batch differences in the extent of flagellar removal and the residual motility of the cells. Repeated mechanical treatment may also impose variable mechanical stress on the cells, contributing to the relatively large variability observed among independent experiments.}

\subsection{3D real-time tracking technology}

3D real-time tracking technology is used to record the three-dimensional swimming trajectory of individual \textit{V.~fischeri} for a long time. The cells are viewed by using an inverted microscope model Nikon ECLIPSE Ti2-U with PX.EDGE 5.5-sCMOS camera.A two-dimensional displacement stage is used to track the movement of cells in the direction of parallel glass plates. The piezoelectric displacement stage (Pi) is used to record the movement of the cells in the vertical direction and control the position of the focal plane to ensure that the cells are always in view. Through the algorithm written by C++ language, the images are obtained at 60fps, 300X300 pixels. And the data of the two-dimensional and the piezoelectric displacement stage are recorded at the same time. The real-time three-dimensional motion trajectory of the cell is finally obtained in the specified time. As shown in Fig.~\ref{speed}B, each cell is continuously tracked for a duration of 10~s, allowing the entire push/pull--wrap--push/pull swimming cycle of each cell to be clearly resolved.

{Prior to swimming-mode identification, the three-dimensional trajectories were smoothed to reduce measurement noise. The instantaneous swimming direction was determined from the displacement vectors between consecutive frames, and the change in swimming direction was calculated from the angle between successive displacement vectors. The onset of the wrap mode was identified when the directional change between two consecutive frames exceeded $10^\circ$. The cell was considered to remain in the wrap mode while its instantaneous swimming speed was below 30\% of the maximum swimming speed within the same trajectory. The duration of each wrap event was determined from the time interval over which this speed criterion was satisfied. In contrast, trajectory segments showing relatively straight motion, with directional changes below $10^\circ$ and relatively high swimming speeds, were classified as the push/pull mode.}

\subsection{Diffusion-based encounter simulation}

A three-dimensional Brownian dynamics simulation is performed in MATLAB to quantify encounters between a motile \textit{V.~fischeri} cell and freely diffusing signal molecules. One bacterium and 10,000 signal molecules are initially distributed at uniformly random positions within a cubic domain with a side length of $300~\mu\mathrm{m}$. The motion of the bacterium is represented by the experimentally determined effective diffusion coefficient, $D_{\mathrm{cell}}$, for each viscosity and fluid condition. The diffusion coefficient of the signal molecules, $D_{\mathrm{signal}}$, is calculated using the Stokes--Einstein relation: $D_{\mathrm{signal}} = k_{\mathrm{B}}T / (6\pi\eta r_{\mathrm{h}})$, where $k_{\mathrm{B}}$ is the Boltzmann constant, $T=289.15~\mathrm{K}$, $\eta$ is the dynamic viscosity. The 3-oxo-C6-HSL molecule is approximated as a spherical particle
with a density comparable to that of the surrounding liquid. Based
on its molecular weight of $213.23~\mathrm{g/mol}$, its equivalent
spherical radius is estimated to be $0.44~\mathrm{nm}$ and used as
an approximation of the hydrodynamic radius in the Stokes--Einstein
relation~\cite{Erickson_2009}. Reflecting boundary conditions are applied to all six surfaces of the simulation domain. Simulations are conducted with a time step of $\Delta t=0.01~\mathrm{s}$ for a total duration of $200~\mathrm{s}$. An encounter is registered when the center-to-center distance between the bacterium and a signal molecule is less than or equal to $0.10~\mu\mathrm{m}$. Each signal molecule is counted only once, even if it subsequently encounters the bacterium multiple times. The mean number of distinct signal molecules encountered by the bacterium is used as a measure of signal-searching efficiency. For comparison across fluid conditions, the mean encounter count is normalized by the corresponding value obtained at a viscosity of $1~\mathrm{cP}$.

\endgroup

\section{Acknowledgments}
We thank Yu Cheng and Jin Zhu for the comments and discussions on the paper content. Our work is funded by National Natural Science Foundation of China (NSFC), Grant No. 12202275, and Shanghai Jiao Tong University Explore X Grant.


\end{document}


\maketitle

\begin{figure}[!htbp]
    \centering
    \includegraphics[width=1\linewidth]{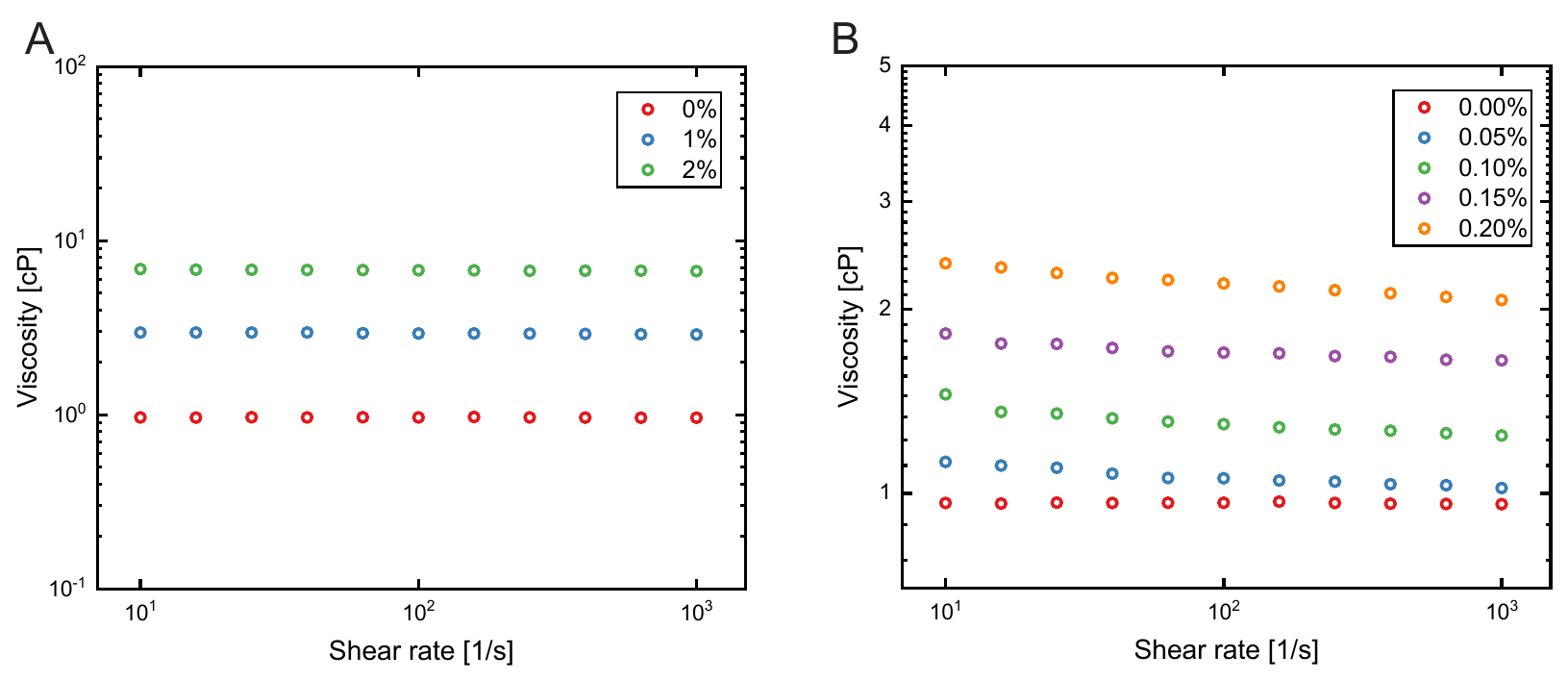}\\
	\caption{Rheological characterization of polymer solutions used in this study.
Shear viscosity as a function of shear rate for (A) PVP360k solutions, exhibiting Newtonian behavior, and (B) Methocel solutions.}
	\label{4-3}
\end{figure}

\begin{figure}
\centering
\includegraphics[width=0.55\linewidth]{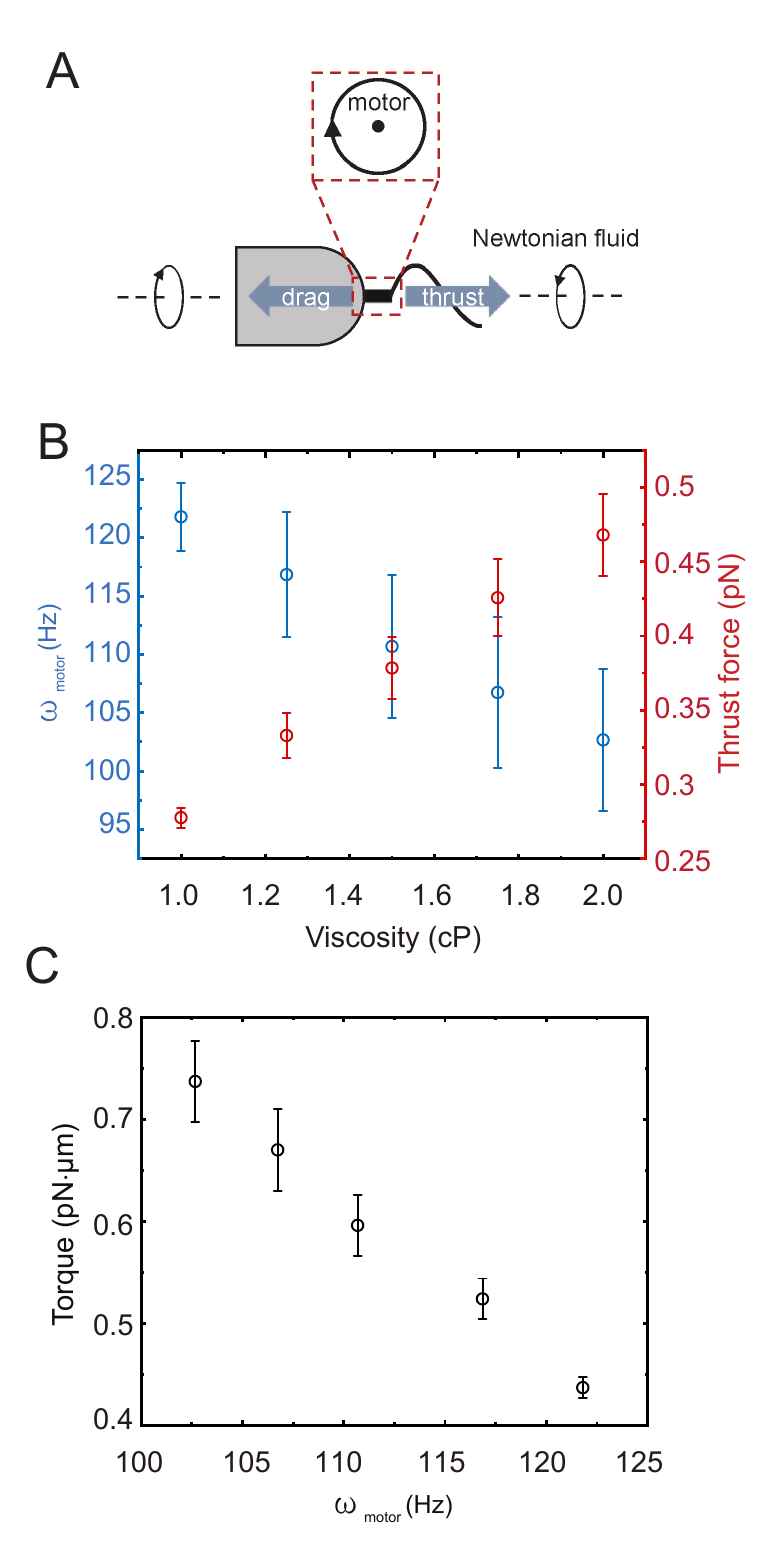}
\caption{
{Hydrodynamic interpretation of viscosity-dependent swimming in \textit{V.~fischeri} based on resistive force theory (RFT).
(A) Schematic illustration of the force balance during bacterial swimming in a Newtonian fluid (PVP360k-containing buffer). Propulsion is governed by the balance between motor-generated thrust and viscous drag.
(B) Viscosity dependence of the motor rotation frequency $\omega_{\mathrm{motor}}$ (left axis) and the corresponding propulsive thrust force (right axis), as predicted by RFT. Error bars represent the standard error of the motor rotation frequency and propulsive thrust force.
(C) Motor torque as a function of the motor rotation frequency $\omega_{\mathrm{motor}}$, calculated using RFT with the geometric parameters listed in Table~\ref{tab:parameters}. Error bars represent the standard error of the calculated motor torque.} \label{RFT}}
\end{figure}

\begin{figure}[!htbp]
    \centering
    \includegraphics[width=0.7\linewidth]{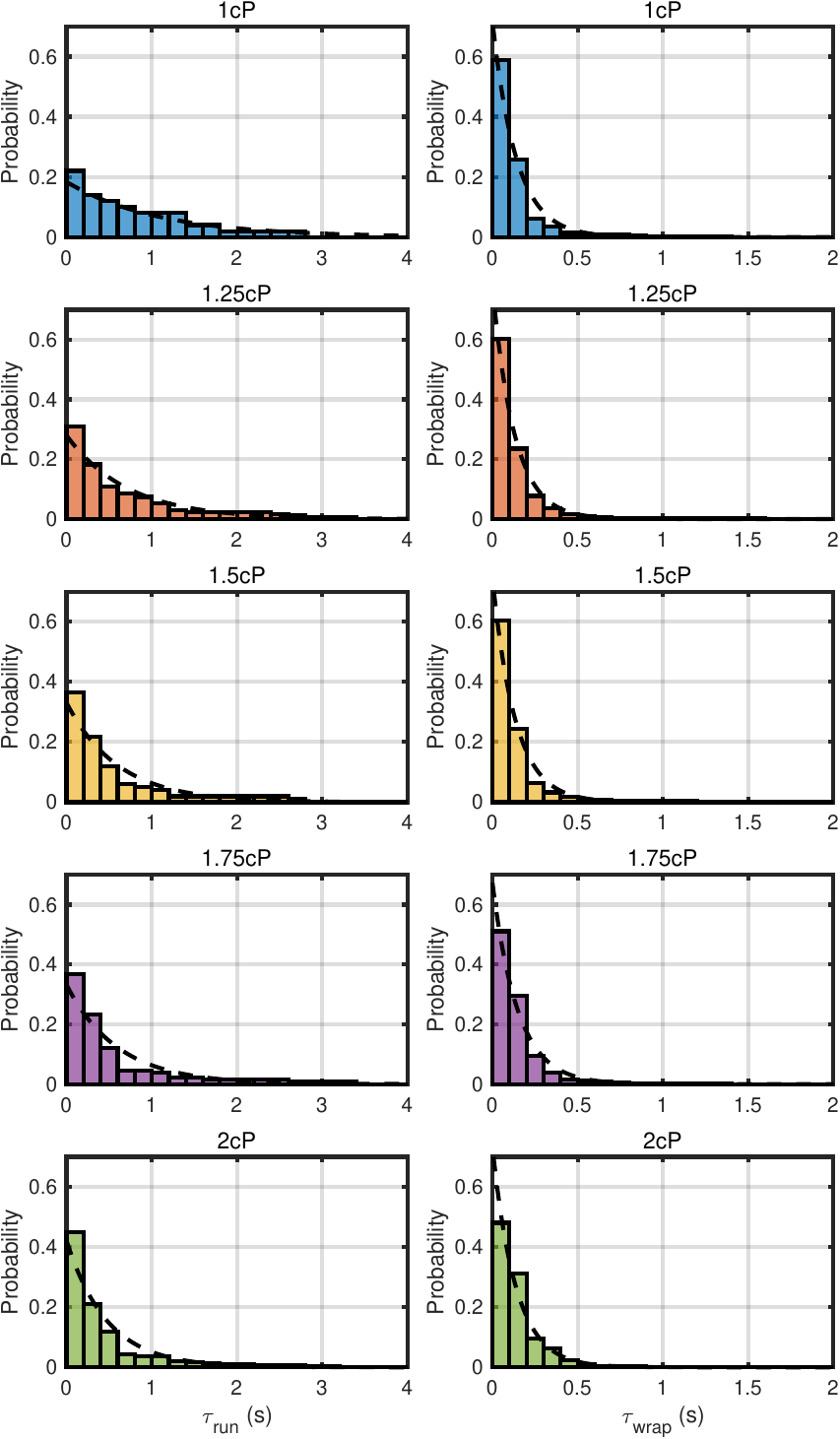}\\
	\caption{Probability distributions of run durations $\tau_{\mathrm{run}}$ (left) and wrap durations $\tau_{\mathrm{wrap}}$ (right) for \textit{V.~fischeri} swimming in PVP360k solutions at different viscosities. Dashed lines indicate exponential fits, showing that both run and wrap durations follow exponential statistics.}
	\label{4-11}
\end{figure}

\begin{figure}[!htbp]
    \centering
    \includegraphics[width=0.7\linewidth]{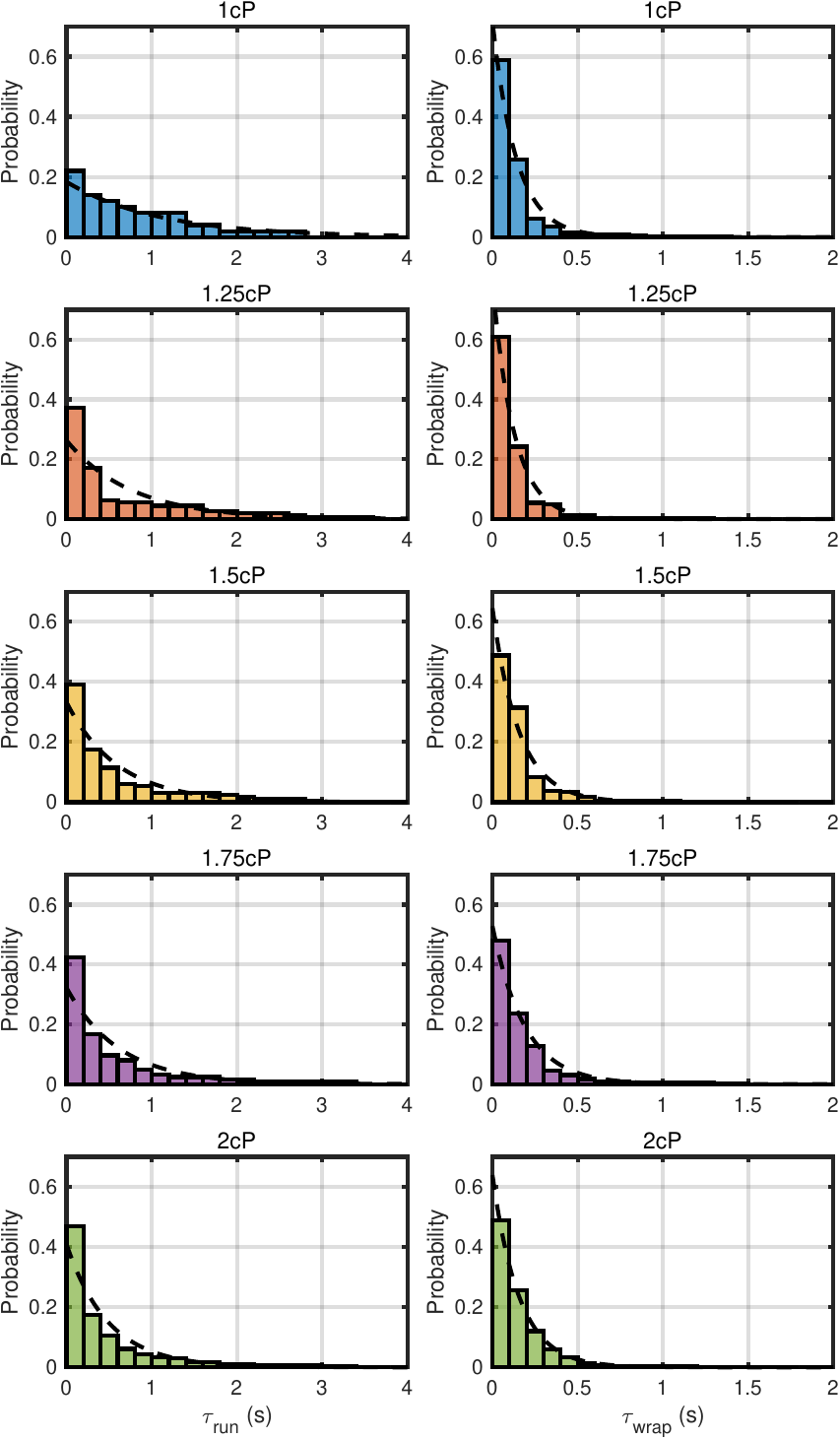}\\
	\caption{Probability distributions of run durations $\tau_{\mathrm{run}}$ (left) and wrap durations $\tau_{\mathrm{wrap}}$ (right) for \textit{V.~fischeri} swimming in Methocel solutions at different viscosities. Dashed lines represent exponential fits, indicating exponential distributions of both run and wrap durations.}
	\label{4-12}
\end{figure}

\begingroup

\section{Fluid rheological properties}

To investigate how fluid viscosity and rheological properties influence quorum-sensing (QS)--activated bioluminescence in \textit{V.~fischeri}, we systematically vary the physical properties of the surrounding media. Two representative polymers are employed to separately examine the effects of increased viscosity and those associated with non-Newtonian rheology. The rheological properties of these polymer solutions are characterized over a broad range of shear rates. As shown in Fig.~\ref{4-3}A, PVP360k solutions exhibit an approximately shear-rate–independent viscosity across the measured range, indicating that the fluid retains a nearly constant viscosity even at higher polymer concentrations. In contrast, the low-concentration Methocel solutions exhibit a weak but systematic shear-rate dependence, with viscosity decreasing slightly as the shear rate increases within the measured range (Fig.~\ref{4-3}B). For selected concentrations, PVP360k and Methocel solutions are adjusted to have comparable viscosities in the low-shear regime relevant to bacteria swimming, allowing a direct comparison between Newtonian and non-Newtonian fluids with similar viscous resistance but different rheological behavior. {Previous microrheological studies have reported measurable viscoelasticity in dilute Methocel solutions at concentrations comparable to those used here~\cite{Qu2023Microrheology}. In the present study, however, the rheological characterization is limited to bulk shear viscosity as a function of shear rate. Other rheological quantities, including relaxation time, normal stress, and extensional rheology, were not independently measured.}

\section{\label{sec:level1}Analysis of the resistive force theory of \textit{V.~fischeri}}

In Newtonian fluids, the rotation of the flagella generates thrust that balances the hydrodynamic resistance on the cell body, leading to stable forward motion (shown in Fig.~\ref{RFT}A). Resistive Force Theory (RFT) is used to analyze the motor drive characteristics~\cite{Qu_et_al_2018}. In RFT, the dynamical characteristics of the cell are determined by the geometric properties of the cell body and flagella, and the geometric parameters used in the calculations are summarized in Table~\ref{tab:parameters}. As shown in Fig.~\ref{RFT}B, the motor torque decreases with increasing rotation frequency, indicating a non-constant torque-speed characteristic. This hydrodynamic response leads to a viscosity-dependent reduction in the motor rotation rate, while the corresponding propulsive thrust force increases moderately with viscosity due to enhanced viscous resistance (Fig.~\ref{RFT}C). 

{
Following Martinez et al.~\cite{Martinez2014Flagellated}, the characteristic local shear rate near a rotating helical flagellum can be estimated as
\begin{equation}
\dot{\gamma}_f \sim \frac{\omega D}{d},
\end{equation}
where $\omega$ is the angular rotation frequency, $D$ is the flagellar helix diameter, and $d$ is the filament thickness. Using the geometrical parameters adopted in our RFT calculation ($D = 0.54~\mu\mathrm{m}$ and $d = 0.10~\mu\mathrm{m}$) and the calculated motor rotation frequencies of approximately 103--122~Hz over the viscosity range of 1--2~cP, the characteristic local shear rate is estimated to be approximately
\begin{equation}
\dot{\gamma}_f \approx (3.5\text{--}4.1)\times10^3~\mathrm{s^{-1}}.
\end{equation}
This value represents an order-of-magnitude estimate rather than a direct measurement of the near-flagellum shear rate.}

\begin{table}[t]
\caption{Typical geometric parameters used in cell swimming calculations~\cite{Kinosita_et_al_2018}.}
\label{tab:parameters}
\centering
\begin{tabular}{c l c}
\toprule
Symbol & Description & Value \\
\midrule
$a$   & Cell length & $1.70~\mu\mathrm{m}$ \\
$b$   & Cell width  & $0.79~\mu\mathrm{m}$ \\
$l$   & Flagellum length & $4.20~\mu\mathrm{m}$ \\
$p$   & Flagellum pitch & $2.40~\mu\mathrm{m}$ \\
$R$   & Flagellum helix radius & $0.27~\mu\mathrm{m}$ \\
$r_0$ & Flagellum filament radius & $0.05~\mu\mathrm{m}$ \\
\bottomrule
\end{tabular}
\end{table}

\section{\label{sec:level1}Derivation of the effective diffusion coefficient}

Within a time window $t$, the cell undergoes approximately $t/\tau_{\mathrm{run}}$ reorientation events, each contributing a factor $\langle \cos\theta\rangle$ to the decay of orientational memory. After $t/\tau_{\mathrm{run}}$ such events, the orientation autocorrelation function can be written as:
\begin{equation}
    \left\langle\mathbf{e}(0)\cdot\mathbf{e}(t)\right\rangle=\exp\left[-\frac{t}{\tilde{\tau}}\right],\quad\tilde{\tau}=\frac{\tau_{\mathrm{run}}}{1-\langle\cos\theta\rangle}
\end{equation}

\begingroup

The corresponding velocity autocorrelation function then takes the form~\cite{Kubo_1957}:
\begin{equation}
    C_{\mathbf{}}(t_1,t_2)=v^2\exp\left(-\frac{|t_2-t_1|}{\tilde{\tau}}\right),
\end{equation}
The mean square displacement can be obtained from the velocity autocorrelation function as:
\begin{equation}
    \langle[\mathbf{r}(t)-\mathbf{r}(0)]^{2}\rangle_{\mathrm{}}=2v^{2}\tilde{\tau}^{2}\left(\frac{t}{\tilde{\tau}}-1+e^{-t/\tilde{\tau}}\right),
\end{equation}
In a three-dimensional environment, when t is large enough, the ordinary diffusion satisfies~\cite{Codling_et_al_2008}:
\begin{equation}
    \langle\Delta r^2(t)\rangle\approx6Dt,\quad\langle[\mathbf{r}(t)-\mathbf{r}(0)]^{2}\rangle_{\mathrm{}}\sim6Dt
\end{equation}
From this expression, the effective diffusion coefficient is obtained as~\cite{Lovely_Dahlquist_1975,Taktikos_et_al_2013}:
\begin{equation}
    D=\frac{v^2\tau_{\mathrm{run}}}{3\left(1-\langle\cos\theta\rangle\right)}.
    \label{D}
\end{equation}
\endgroup

Analysis of the trajectory statistics shows that the distribution of run and wrap durations is well described by an exponential form (Fig.~\ref{4-11} and Fig.~\ref{4-12}), supporting a Poisson description of reorientation events.

 \endgroup

\clearpage

\begin{figure}
\centering
\includegraphics[width=1\linewidth]{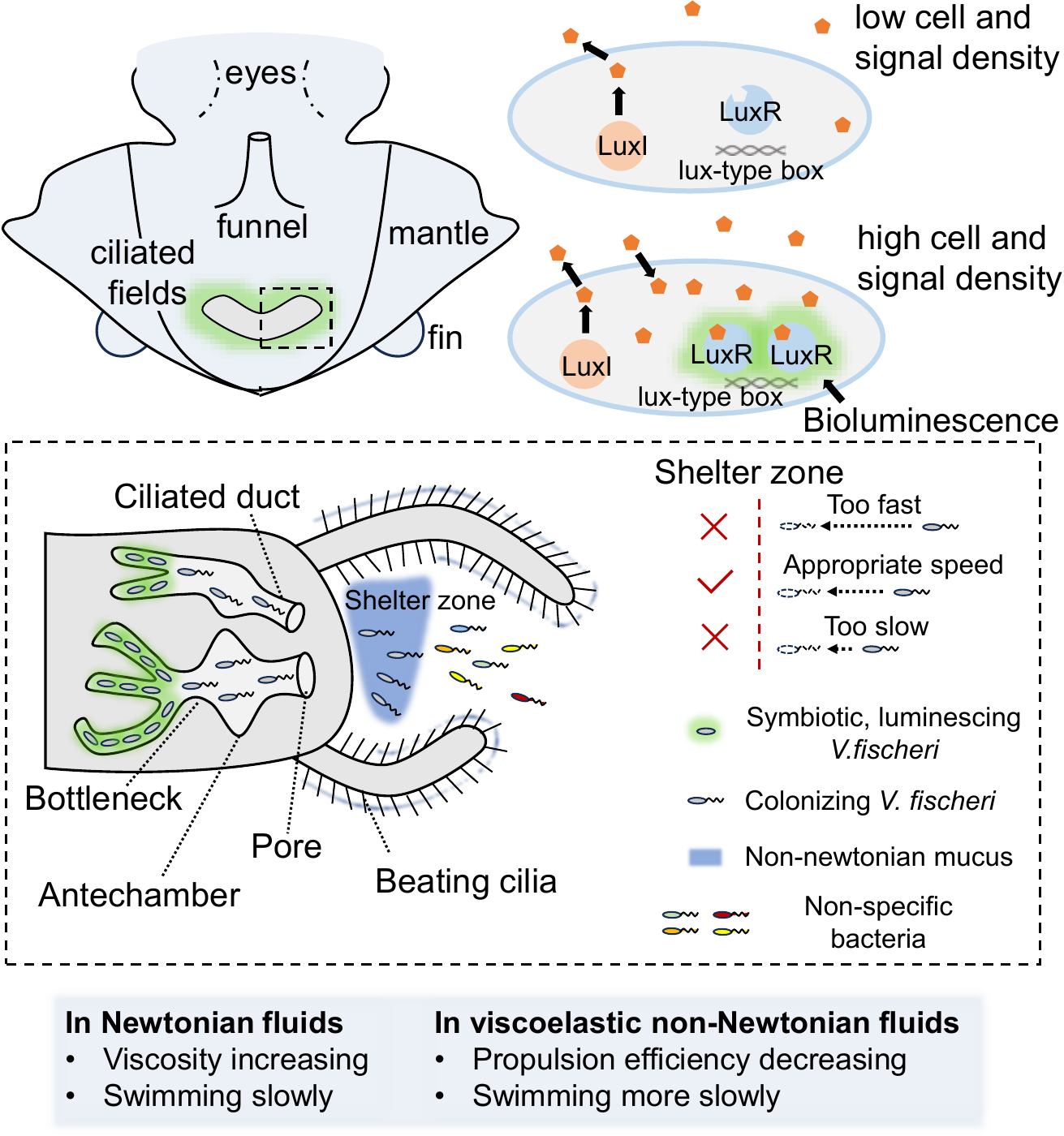}
\caption{Schematic of quorum-sensing (QS) regulation in \textit{V.~fischeri}. 
At low cell and signal density, LuxR remains inactive; accumulation of autoinducer molecules at high density activates LuxR binding to the lux-type box, triggering bioluminescence. 
Conceptual model of \textit{V. fischeri} colonization in the light organ of the Hawaiian bobtail squid (\textit{Euprymna scolopes}). 
Beating cilia transport bacteria toward the light-organ ducts while secreting non-Newtonian mucus. 
Bacteria with swimming speeds that are too fast or too slow are inefficiently captured, whereas an appropriate swimming speed promotes entry into the shelter zone and successful colonization.
Increased viscosity and viscoelasticity reduce propulsion efficiency, leading to slower swimming in non-Newtonian fluids, consistent with the experimentally observed motility reduction and its impact on QS-mediated bioluminescence.
Illustration of bacterial space exploration characterized by different effective diffusion coefficients. 
Larger effective diffusion coefficients correspond to more efficient exploration and higher encounter rates between cells and signal molecules. 
\label{overview}}
\end{figure}

\begin{figure}[!htbp]
    \centering
    \includegraphics[width=0.8\linewidth]{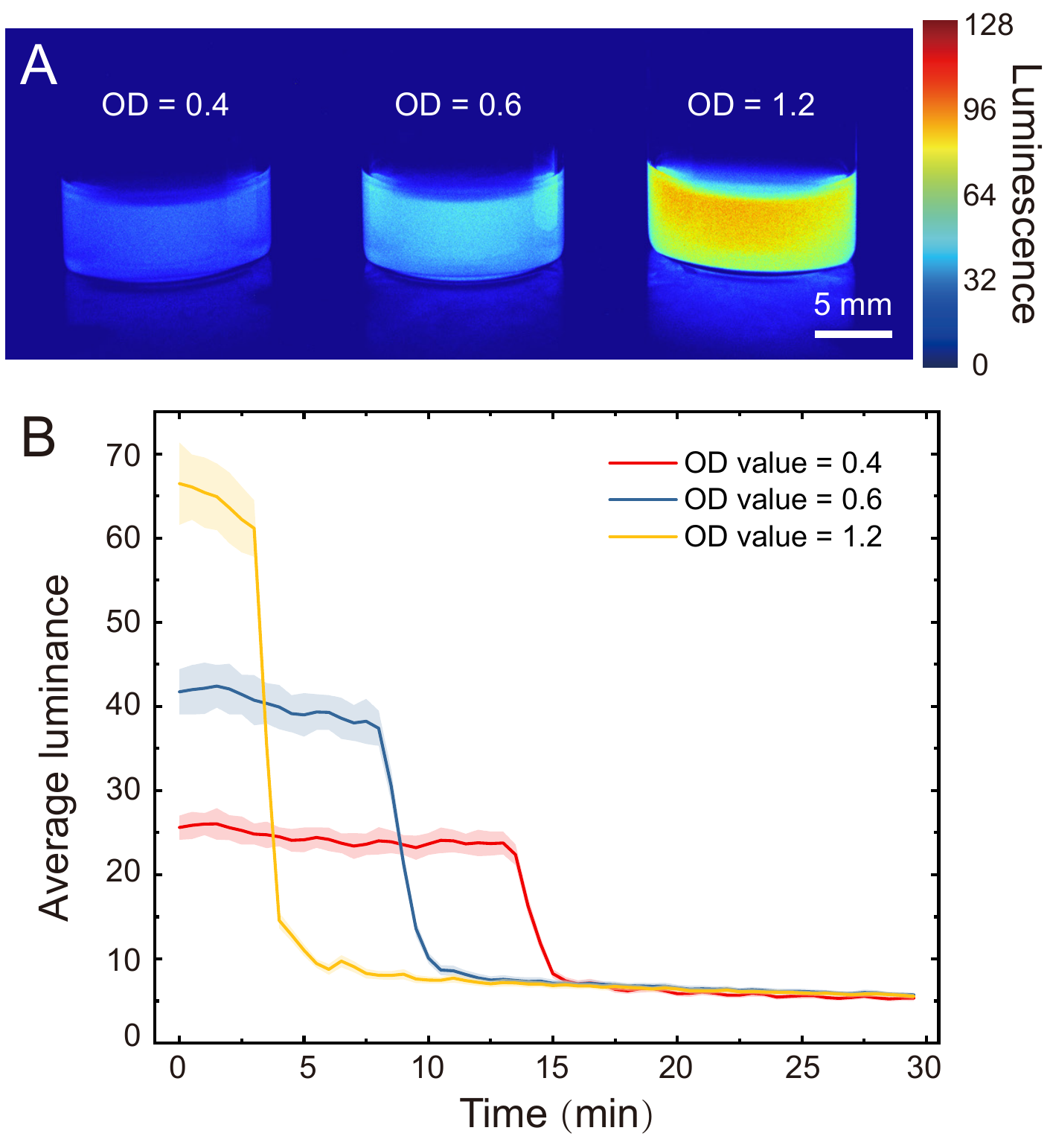}\\
	\caption{Dependence of quorum-sensing–activated bioluminescence on cell concentration.
(A) Representative images of \textit{V.~fischeri} suspensions at different OD$_{600}$ values. 
(B) Time evolution of the average luminescence, illustrating a positive correlation between peak intensity and cell concentration, accompanied by a reduced luminescence duration at higher densities.}
	\label{4-OD value}
\end{figure}

\clearpage 
\bibliographystyle{unsrt}
\bibliography{ref}

\clearpage
\paragraph{Caption for Movie S1.}
\textbf{3D real-time tracking of \textit{V. fischeri} in a motility buffer with a viscosity of 1 cP}

\paragraph{Caption for Movie S2.}
\textbf{3D real-time tracking of \textit{V. fischeri} in a PVP360k containing motility buffer with a viscosity of 1.5 cP}

\paragraph{Caption for Movie S3.}
\textbf{3D real-time tracking of \textit{V. fischeri} in a PVP360k containing motility buffer with a viscosity of 2 cP}

\paragraph{Caption for Movie S4.}
\textbf{3D real-time tracking of \textit{V. fischeri} in a Methocel containing motility buffer with a viscosity of 1.5 cP}

\paragraph{Caption for Movie S5.}
\textbf{3D real-time tracking of \textit{V. fischeri} in a Methocel containing motility buffer with a viscosity of 2 cP}

\paragraph{Caption for Movie S6.}
\textbf{Quorum-sensing activated bioluminescence in \textit{V.~fischeri}}

\paragraph{Caption for Movie S7.}
\textbf{\textit{V.~fischeri} cells without flagella-shearing treatment}

\paragraph{Caption for Movie S8.}
\textbf{\textit{V.~fischeri} cells subjected to flagella-shearing treatment}